\documentclass[showpacs,twocolumn,prl]{revtex4-1}
\usepackage[utf8]{inputenc}
\usepackage{graphicx}
\usepackage{color}
\usepackage{appendix}
\usepackage{amsmath}
\usepackage{amssymb}
\usepackage{subfigure}
\usepackage{epsfig}
\usepackage{ulem}

\newcommand{\om}{\Omega}
\newcommand{\omdag}{{\Omega}^{\dagger}}
\newcommand{\omeq}{\Omega_{\rm eq}}

\newcommand{\delom}{\delta\Omega}

\newcommand{\F}{\boldsymbol{F}}

\newcommand{\pp}{\boldsymbol{p}}

\newcommand{\p}{\boldsymbol{p}}

\newcommand{\rr}{\boldsymbol{r}}
\newcommand{\nab}{\boldsymbol{\nabla}}
\newcommand{\xxi}{\boldsymbol{\xi}}

\newcommand{\eeta}{\boldsymbol{\eta}}
\newcommand{\R}{\boldsymbol{R}}

\begin{document}

\title{Brownian systems with spatially inhomogeneous activity}

\author{A.~Sharma}
\author{J.M.~Brader}
\affiliation{Department of Physics, University of Fribourg, CH-1700 Fribourg, Switzerland}


\begin{abstract}
We generalize the Green-Kubo approach, previously applied to bulk systems 
of spherically symmetric active particles [J. Chem. Phys. {\bf 145}, 161101 (2016)], 
to include spatially inhomogeneous activity. 
The method is applied to predict the spatial dependence of the average orientation 
per particle and the density. 
The average orientation is given by an integral over the self-part of the van Hove function 
and a simple Gaussian approximation to this quantity yields an accurate analytical 
expression.  
Taking this analytical result as input to a dynamic density functional theory approximates 
the spatial dependence of the density in good agreement with simulation data. 
All theoretical predictions are validated using Brownian dynamics simulations.

  
%
%
%
%
\end{abstract}


\maketitle


Active Brownian particles (ABPs) are intrinsically nonequilibrium systems, constantly 
driven out of equilibrium by consuming energy from the local environment. 
Nonequilibrium statistical mechanics aims to calculate from the microscopic 
dynamics the relevant average quantities. 
However, this presents a difficult theoretical 
problem, even for simple model systems. 
One way to make systematic progress is to restrict attention to the linear response 
regime, where the problem becomes tractable. 
Within linear response, formally exact results for system averages can be obtained by 
integrating the time-correlation functions of the corresponding passive 
(equilibrium) 
system~\cite{green1952markoff,kubo1957statistical}.

We have recently applied the linear response (Green-Kubo) approach to a homogeneous system of 
ABPs~\cite{sharma2016communication}, focusing our attention on calculation of the 
average swim speed~\cite{cates2014motility,krinninger2016nonequilibrium}, a central 
quantity appearing in coarse-grained theories of active matter.
We have demonstrated that the average swim speed, which describes how the motion of each particle is 
obstructed by its neighbours, can be obtained from a history integral over the equilibrium 
autocorrelation of tagged-particle force fluctuations~\cite{sharma2016communication}.  The theory was tested using active Brownian dynamics simulations and 
provides a solid basis for the development of first-principles theoretical approaches.

Bulk systems have been the focus of much attention, due largely to the 
phenomenon of motility-induced phase separation, and several experimental studies of bulk 
ABPs have been performed~\cite{howse2007self,jiang2010active,palacci2013living,gomez2016dynamics,palacci2010colloidal}.
However, there exist synthetic~\cite{saha2014clusters,hong2007chemotaxis,magiera2015trapping,stenhammar2016light} and living systems~\cite{berg1972chemotaxis,jekely2008mechanism,hoff2009prokaryotic,jikeli2015sperm} 
for which the propulsion strength is not a global constant, but dependent on the spatial location 
of the particles. For example, in a recent experimental study of synthetic microswimmers~\cite{lozano2016phototaxis} position-dependent motility was implemented using an inhomogenous laser field, resulting in phototaxis. Position-dependent activity also features in the energy depot model~\cite{schweitzer1998complex}.

Motivated by these considerations, we consider in this paper active systems for which the 
particle propulsion-speed varies in space. 
Our aim is to extend the Green-Kubo approach, previously applied in 
bulk~\cite{sharma2016communication}, to treat systems with inhomogeneous activity, 
thus providing a first-principles theoretical route to addressing more realistic situations.
The physical observables to be considered are the average orientation, which features 
prominently in studies of inhomogenous systems~\cite{enculescu2011active,lozano2016phototaxis}, 
and the density.
We will show that both of these quantities become inhomogeneous in the presence of 
spatially varying activity and we calculate these explicitly for a simple test case. 
Our formalism also makes clear that an external activity field affects the system 
in a qualitatively different way than an external potential field; the former 
generates a linear response for the average orientation, whereas the latter does not.





In our Green-Kubo approach we employ a variation of the integration-through-transients approach, originally developed for treating interacting Brownian particles subject to external flow 
\cite{morriss2007statistical,fuchs2002theory,fuchs2005integration,brader2012first}. 
We find that the average orientation is proportional to
the local gradient of the activity field, even in the absence 
of a one-body torque. 
The relevant autocorrelation function is the well-known self part of the van Hove 
function~\cite{hansen1990theory}, which can be very well approximated by a Gaussian. 
Within the Gaussian approximation a simple and accurate analytical expression 
can be obtained for the average orientation.
Taking this as input to a Dynamic Density Functional theory we 
then proceed to develop a closed theory for the inhomogenous density. Our 
predictions are tested against data from 
Brownian dynamics simulations.


We consider a three dimensional system of $N$ active, interacting, spherical Brownian particles 
with coordinate $\rr_i$ and orientation specified by an embedded unit vector $\p_i$. 
A space- and time-dependent self-propulsion of speed $v_0(\rr_i,t)$ 
acts in the direction of orientation and $i$ labels the particle. 
Omitting hydrodynamic interactions the motion can be modelled by the Langevin equations
\begin{align}\label{full_langevin}
&\!\!\!\!\!\!\dot{\rr}_i = v_0(\rr_i,t)\,\p_i  + \gamma^{-1}\F_i + \xxi_i\;\;,
\;\;\;
\dot{\p}_i = \eeta_i\times\p_i \,,
\end{align}
where $\gamma$ is the friction coefficient and the force on particle $i$ is generated from the total interparticle interaction energy according to $\F_i\!=\!-\nabla_i U_N$. 
For clarity of presentation we do not include an external potential field.
The stochastic vectors $\xxi_i(t)$ and $\eeta_i(t)$ are Gaussian distributed with zero mean and 
have time correlations	
$\langle\xxi_i(t)\xxi_j(t')\rangle=2D_t\boldsymbol{1}\delta_{ij}\delta(t-t')$ and 
$\langle\eeta_i(t)\eeta_j(t')\rangle=2D_r\boldsymbol{1}\delta_{ij}\delta(t-t')$. 
The translational and rotational diffusion coefficients, $D_t$ and $D_r$, are treated 
as independent parameters. 
Note that in the second of Eqs.~\eqref{full_langevin} the orientation vector does not 
couple to the activity field and hence no direct torque acts on the particle due to the 
position-dependent activity.


It follows exactly from \eqref{full_langevin} that the  
joint probability distribution, $\!P(\rr^N\!\!,\p^N\!\!,t)$, evolves according to 
\cite{gardiner1985handbook}
\begin{align}\label{smol_eq}
\frac{\partial P(t)}{\partial t} = \om_{\rm a}(t) P(t),
\end{align}
where $\om_{\rm a}$ is the time-evolution operator. We have used $\!P(\rr^N\!\!,\p^N\!\!,t) \equiv P(t)$, and $\om_{\rm a}(\rr^N\!\!,\p^N\!\!,t) \equiv \om_{\rm a}(t)$ to keep the notation compact.
The time-evolution operator can be split into a sum of two terms, $\om_{\rm a}(t)=\omeq+\delom_{\rm a}(t)$, 
where the equilibrium contribution is given by 
\begin{align}\label{smol_op_eq}
\omeq = \sum_{i=1}^{N} \nab_{i}\!\cdot\!
\big[
D_\text{t}\!\left(\nab_{i}\! - \beta\F_i\right) 
\big] \!+\! D_\text{r}\R_i^2, 
\end{align}
with rotation operator $\R\!=\!\p\times\!\nabla_{\!\p}$ \cite{morse1953methods} and $\beta = 1/(k_{\rm B}T)$. 
Using $\omeq P_{\rm eq}=0$, we obtain a formal solution for the nonequilibrium distribution
\begin{align}\label{P_eom}   
P(t) = P_{\rm eq} \,+\, \int_{-\infty}^{t}\!dt' 
\,e_{+}^{\int_{t'}^{t}ds\,\om_{\rm a}(s)}
\delom_{\rm a}(t)P_{\rm eq}\,,
\end{align}
where $e_{+}$ is a positively ordered exponential function~\cite{brader2012first}. 
The active part of the dynamics is described by the operator $\delom_{\rm a} = -\sum_i \nab_{i}\!\cdot (v_0(\rr_i,t)\p_i)$. The action of this operator on $P_{\rm eq}$ yields 
$\delom_{\rm a}(t)P_{\rm eq} =-P_{\rm eq}\left(K(t) + V(t)\right)$,
where we have defined the quantities $K(t)$ and $V(t)$ as
\begin{align}
K(t) &= \sum_{i=1}^{N} v_0(\rr_i,t)\,\p_i \cdot \beta\F_i \label{KP},\\
V(t) &= \sum_{i=1}^{N} \p_i\cdot \nabla_i v_0(\rr_i,t)  
\label{VP}
\end{align}

%
%
We obtain from Eqs.~\eqref{P_eom},~\eqref{KP} and~\eqref{VP} an exact expression for 
the nonequilibrium average of a test function $f \equiv f(\rr^N\!\!,\p^N\!)$ as
\begin{align}\label{faverage}
\langle f \rangle(t) = \langle f \rangle_{\rm eq} - \int_{-\infty}^{t}\!\!dt\,
\langle G(t') e_{-}^{\int_{t'}^{t}ds\,\omdag_{\rm a}(s)}f \rangle_{\rm eq}, 
\end{align}
where we have defined $G(t) = K(t) + V(t)$ and the adjoint operator is given by 
$\omdag_{\rm a}(t)=\omdag_{\rm eq}-\delom_{\rm a}(t)$, where $\omdag_{\rm eq}=\sum_{i} 
D_\text{t}\!\left(\nab_{i}\! + \beta\F_i\right) 
\!\cdot\!\nab_{i} \!+\! D_\text{r}\R_i^2$. 
The integrand appearing in Eq.~\eqref{faverage} involves the equilibrium correlation 
between $G$ at time $t'$ and the observable $f$ which evolves from $t'$ to $t$ according 
to the full dynamics. 
From here onwards we will consider only the linear response, 
obtained by replacing the full time-evolution operator in 
\eqref{faverage} by the time-independent equilibrium adjoint operator.

We will focus first on calculating to linear order in activity the average orientation per 
particle, defined as
\begin{align}
\p(\rr) = \frac{\left\langle \sum_i \delta(\rr - \rr_i)\p_i\right\rangle}{\rho(\rr)},
\label{averageP}
\end{align}
where $\rho(\rr) = \left\langle \sum_i \delta(\rr - \rr_i)\right\rangle$ is the one-body density. 
We will henceforth assume that the activity does not vary in time, 
$v_0(\rr_i,t)=v_0(\rr_i)$; generalization to time-dependent situations is straightforward.

Our first remark is that any observable independent of the particle 
orientation vectors, $\pp_i$, does not admit a linear response in $v_0(\rr)$. 
As the function $G(t)$ appearing in Eq.~\eqref{faverage} is linear in $\pp_i$ the 
angular integrals in the equilibrium average will yield zero by symmetry. 
Only odd functions of $\pp_i$ will generate a nonzero linear-response.   
As the density is independent of $\pp_i$ we can thus replace $\rho(\rr)$ in 
Eq.~\eqref{averageP} by the bulk number density $\rho_b$ when working to first order 
in $v_0$.

The average orientation per particle is obtained by using 
Eq.~\eqref{faverage} to evaluate the numerator 
of Eq.~\eqref{averageP} to linear order in $v_0$. 
The equilibrium time evolution operator can be split into translational and rotational 
contributions,  
$\omdag_{\rm eq} = \omdag_{\rm eq,r} +\, \omdag_{\rm eq,tr}$. 
These operators commute (the rotational part 
acts only upon the particle orientation), which allows the angular integrals to be 
evaluated explicitly. 
This yields
\begin{align}
\pp(\rr)\!=
 -\!\!\int_{0}^{\infty}\!\!\!dt\,\frac{e^{-2D_{r}t}}{3\rho_b}\left\langle \sum_{i} \nab_i v_0(\rr_i) e^{\omdag_{\rm eq,tr}t} \delta(\rr - \rr_i)\!\right\rangle_{\rm \!\!eq,s}
\label{pfull}
\end{align}
where $\langle..\rangle_{\rm eq,s}$ denotes 
an average over the spatial degrees of freedom. 
In deriving Eq.~\eqref{pfull} we have employed two results: 
Firstly, the orientation decorrelates according to 
%
$\langle\p_i e^{\omdag_{\rm eq,r}t}\p_j\rangle_{\rm eq,r} 
\!\!= \delta_{ij} \boldsymbol{1} e^{-2D_{r}t}/3$,
%
where the equilibrium average is over rotational 
degrees of freedom (see the supplementary material). 
Secondly, in homogeneous equilibrium the interaction force on a particle is not 
correlated with its position 
\begin{align}\label{zero}
\left\langle \sum_{i}v_0(\rr_i)\F_i \cdot e^{\omdag_{\rm eq,tr}t} 
\delta(\rr - \rr_i)\right\rangle_{\rm eq,s} \!\!\!\!\!= 0.
\end{align}
Using $\nab_i v_0(\rr_i) = \int d\rr \,\delta(\rr - \rr_i)\nab v_0(\rr)$ 
and introducing the self-part of the equilibrium van Hove function~\cite{hansen1990theory}, 
%
$G_{\rm vH}^{\rm s}(|\rr-\rr'|,t) = 
\langle\delta(\rr' - \rr_1)e^{\omdag_{\rm eq,tr}t} \delta(\rr - \rr_1\rangle_{\rm eq,s}$,
%
enables us to express the average orientation in the compact form
\begin{align}
\p(\rr) = -\int_{0}^{\infty}\!dt\,\frac{e^{-2D_{r}t}}{3}\int d\rr'  \nab v_0(\rr') 
G_{\rm vH}^{\rm s}(|\rr-\rr'|,t).
\label{panalytic}
\end{align} 
The average orientation of particles is antiparallel to the gradient of the activity field and 
has a magnitude determined by the equilibrium self van Hove function. 
This result can be rewritten in the alternative form
\begin{align}
\pp(\rr) &= \int_0^{\infty}\!dt\int d\rr' v_0(\rr')\,\chi(|\rr - \rr'|,t),
\label{pexpression}
\end{align}
where the space-time response function, $\chi(|\rr - \rr'|,t)$, 
is given by
\begin{align}
\label{response}
\chi(|\rr - \rr'|,t) =\frac{e^{-2D_{r}t}}{3}\nab G_{\rm vH}^{\rm s}(|\rr-\rr'|,t).
\end{align}
The right hand side of Eq.~\eqref{response} is simply the functional derivative of 
$\pp(\rr)$ with respect to $v_0(\rr')$, evaluated at zero activity. 
Eqs.~\eqref{pexpression} and \eqref{response} are the key linear response 
results of this paper.

Recent experiments have shown that active particles tend to orient in an inhomogenous activity 
field~\cite{bickel2014polarization,lozano2016phototaxis}. 
In these studies, it was identified that the orientation is a consequence of an aligning 
torque acting on the particles. 
Within a first-principles theoretical approach, such a one-body torque would have to be  
explicitly incorporated into the orientational member of the Langevin equations~\eqref{full_langevin}. 
While it is clear that deterministic one-body torques can generate an average orientation, it 
is much less obvious that this will emerge from the present, torque-free Langevin equations as a 
purely statistical phenomenon.  
It is this aspect which is of primary interest in the present study. 
From Eq.~\eqref{panalytic} it is evident that ABPs do indeed tend to orient in an inhomogenous 
activity field, even in the absence of deterministic aligning torques. 
Moreover, our numerical results will demonstrate that this can be a large effect 
within certain parameter ranges. 
A notable feature of the present orientation mechanism is that it is independent of the 
particle diameter, in contrast to the torque-based mechanism identified in 
Refs.~\cite{bickel2014polarization,lozano2016phototaxis}.

We can use the linear response theory to distinguish the orientational response of ABPs to 
inhomogeneous activity from the response to an external potential field (with spatially 
constant activity), 
e.g.~the sedimentation of ABPs under gravity~\cite{enculescu2011active}. 
In the former situation the leading order contribution to $\pp(\rr)$  is linear in 
$v_0$, as is evident from Eq.~\eqref{pexpression}, whereas in the latter situation the leading 
order is quadratic. 
For a system with constant $v_0$ and external potential $v_{\rm ext}(\rr)$ 
the function $G(t)$ appearing in Eq.~\eqref{faverage} is replaced by the time-independent 
function $G=K + D_t\beta\sum_i(\beta {\bf F}_i\cdot\nabla_iv_{\rm ext}(\rr_i) 
+ \nabla^2_iv_{\rm ext}(\rr))$. Use of Eq.~\eqref{faverage} to calculate 
$\pp(\rr)$ to linear order in $v_0$ thus yields three terms, all of which are zero; 
the first vanishes due to Eq.~\eqref{zero} and the others due to the symmetry of the angular 
integrands.

Given the exact linear response result Eq.~\eqref{pexpression}, it is desirable to obtain from 
this a closed theory by approximating the self part of the van Hove function.
A commonly employed approximation is the Gaussian~\cite{hansen1990theory} 
\begin{align}\label{gaussian}
G^{\rm s}_{\rm vH}(\rr,t) 
= \frac{1}{(4\pi D_t t)^{3/2}}e^{-\frac{r^2}{4D_t t}}.
\end{align}
This approximation is known to be accurate, at least for hard-spheres, up to dimensionless 
densities as high as $\rho_b\!=\!0.6$~\cite{hopkins2010van}. 
We will show below that this approximation proves very reliable for calculating the 
average orientation profiles of repulsive ABPs and, for certain choices of $v_0(\rr)$, 
enables evaluation of the integrals in Eq.~\eqref{panalytic} to yield 
an explicit analytical expression. 


As discussed above, 
for reasons of symmetry the density remains unaltered from that in bulk to linear order;  
the first modification is quadratic in $v_0$.
It is in principle straightforward to expand the exact expression in Eq.~\eqref{faverage} to second 
order in activity (noting that $\Omega_{\rm a}^{\dagger}$ is a function 
of $v_0$), however, the formal expression thus generated for the density response 
is difficult to evaluate and yields little insight. We thus
follow an alternative, simpler route to obtain the density approximately.

A coarse-grained expression for the 
density can be obtained by integrating Eq.~\eqref{smol_eq}
over all orientational and all but one translational degrees
of freedom. However, this generates a term involving the two-body density. 
The dynamical density functional theory (DDFT) approximates this unknown term 
using an equilibrium free energy functional. 
In the case of passive particles this is sufficient to yield a closed theory~\cite{hopkins2010van}. 
Applying this procedure to Eq.~\eqref{smol_eq} yields the 
following DDFT 
\begin{align}\label{ddft}
\frac{\partial\rho(\rr,t)}{\partial t}&= 
\nabla\cdot\left(v_0(\rr,t)\rho(\rr,t)\,\pp(\rr,t)\right)
\notag\\
&+
D_t\,\nabla\cdot\left(\rho(\rr,t)
\nabla\frac{\delta \beta\mathcal{F}[\rho]}{\delta\rho(\rr,t)}\right),
\end{align}
where $\mathcal{F}[\rho]$ is the 
Helmholtz free energy functional. 
We observe that Eq.~\eqref{ddft} involves two unknown functions, $\rho(\rr,t)$ 
and $\pp(\rr,t)$, and is thus not closed.  
Our strategy is to use as input to the DDFT the linear response result 
Eq.~\eqref{pexpression} with the Gaussian approximation Eq.~\eqref{gaussian}. 
The resulting equation 
can then be solved self-consistently for the steady-state density. 
Note that the activity 
appears quadratically in the DDFT equation, such that the density is 
independent of the sign of $v_0$.
It now only remains to specify the free energy functional.

The Helmholtz free energy can be split into two 
contributions, $\mathcal{F}=\mathcal{F}_{\rm id}+\mathcal{F}_{\rm ex}$. 
The ideal part is given exactly by
%
$\mathcal{F}_{\rm id}[\rho]=k_BT\!\int d\rr \rho(\rr,t)\left(\log(\Lambda^3\rho(\rr,t))-1\right)$,
%
where $\Lambda$ is the thermal wavelength. The excess part, $\mathcal{F}_{\rm ex}$, 
encodes the interparticle interactions and is, in general, unknown. 
A commonly employed approximation, sufficient for our 
present purposes, is to use a functional Taylor expansion, truncated at second order 
in the density. 
Within this approximation the gradient of the functional derivative entering 
Eq.~\eqref{ddft} is given by
\begin{align}\label{approximation}
\!\!\nabla\frac{\delta\beta\mathcal{F}[\rho]}{\delta\rho(\rr,t)}
= \frac{\nabla\rho(\rr,t)}{\rho(\rr,t)} 
-
\nabla\!\int d\rr' c^{(2)}(|\rr-\rr'|)\tilde\rho(\rr',t),
\end{align}
where $\tilde\rho(\rr,t)=\rho(\rr,t)-\rho_b$ and $c^{(2)}(r)$ is the 
bulk direct correlation function, easily obtained from standard
liquid state integral equation theory. We employ here the Percus-Yevick 
integral equation~\cite{hansen1990theory}. 

Equations \eqref{pexpression}, \eqref{response} 
and \eqref{gaussian} enable calculation of the average orientation 
per particle. Equations \eqref{ddft}-\eqref{approximation} take this as input and yield 
the density. 
We will next present numerical results and compare these 
with data from active Brownian dynamics simulation. The simulations are performed on a three-dimensional system of $N=500$ particles 
interacting via the pair-potential 
$\beta u(r) = 4\varepsilon((\sigma/r)^{12} - (\sigma/r)^6)$, 
where $\sigma$ sets the length scale and we set $\varepsilon=1$. 
The potential is truncated at its minimum, $r=2^{1/6}\sigma$ to yield a softly repulsive 
interaction. 
The system size $L$ is determined as $L\!=\!(N/\rho_{\rm b})^{1/3}$ in order to obtain the desired 
density. Periodic boundary conditions are applied in all three directions. 
The integration time step is fixed to $dt\!=\!10^{-5} \tau_B$ where $\tau_B\!=\!d^2/D_t$ is 
the time-scale 
of translational diffusion. The equation for time evolution of the 
orientation vector (Eq.~\eqref{full_langevin}) is evaluated as an Ito Integral. 
Measurements are made after a time $20\tau_B$ to 
ensure equilibration. We choose the ratio of diffusion coefficients as $D_r/D_t\!=\!25$. 

\begin{figure}[t]
\begin{tabular}{@{}cc@{}}
\includegraphics[width=0.55\columnwidth]{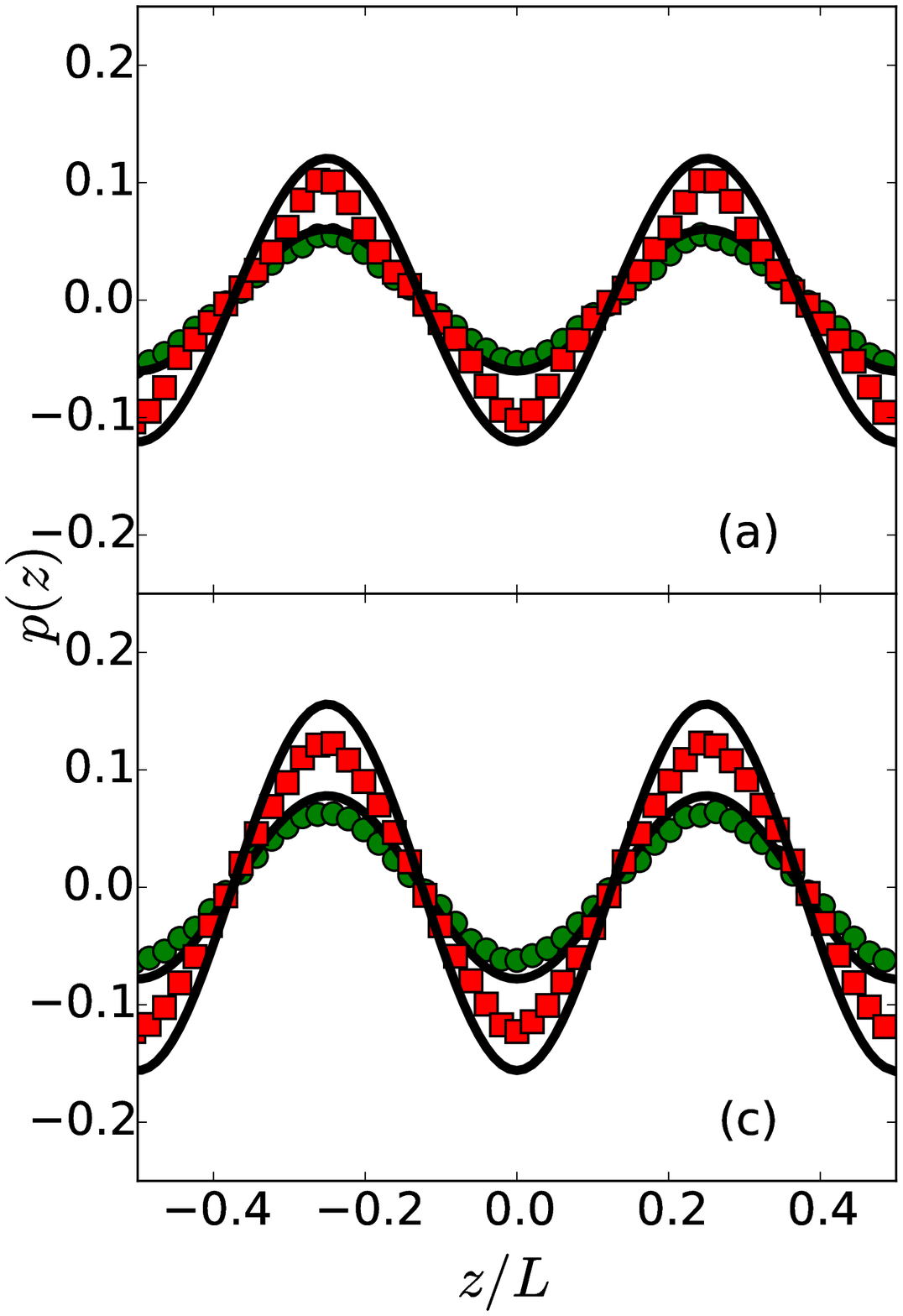} &\hspace{-0.5cm}
\includegraphics[width=0.55\columnwidth]{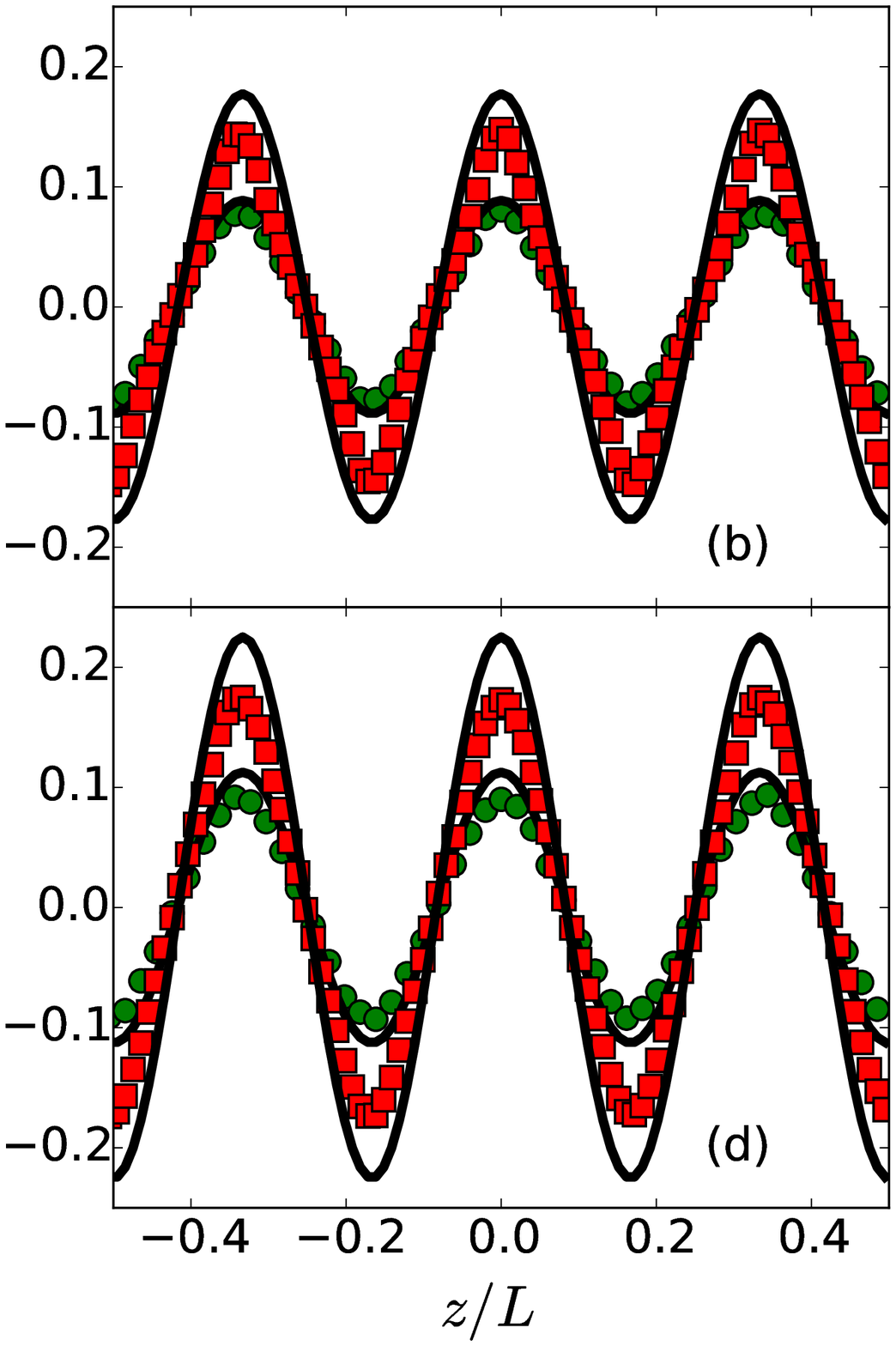}
\end{tabular}
\caption{Average orientation per particle in the $\hat{\boldsymbol{e}}_z$ direction for $\rho_b = 0.2$ in (a) and (b) and $\rho_b = 0.6$ in (c) and (d). The activity field is sinusoidal with amplitude $v_a$ and the angular frequency $\omega = n\omega_0$, where $\omega_0 = 2\pi/L$ and $n$ is a parameter. In (a) and (c), $\omega=2\omega_0$ whereas in (b) and (d), $\omega=3\omega_0$. The circles correspond 
to $v_a = 10$ and the squares to $v_a = 20$. The thick lines corresond to the theoretical prediction of Eq.~\eqref{psinusoidal}. The numerically measured $\pp(z)$ is in very good agreement with the theoretical prediction for low density. At higher density, the theory overestimates the average orientation. Nonlinear deviations are apparent for e.g., in (a) and (b) where higher harmonics contribute to the average orientation per particle.}
\label{fig:orientation}
\end{figure}

As a test case we impose the activity $v_0(z) = v_a \sin(\omega z)$ varying only in the 
$z$-direction, where $v_a$ and $\omega$ are parameters. Inserting this choice and the 
Gaussian approximation into Eq.~\eqref{panalytic} yields a simple theoretical prediction 
for the average orientation per particle
\begin{align}
\pp(z) = -\frac{\nab v_0(z)}{3(2D_r + D_t \omega^2)} = -\boldsymbol{\hat{e}_z} \frac{v_a \omega \cos(\omega z)}{3(2D_r + D_t \omega^2)}.
\label{psinusoidal}
\end{align}
In Fig.~\ref{fig:orientation}, we plot the average orientation per particle
for two different bulk densities. For low density ($\rho_b = 0.2$ in Fig.~\ref{fig:orientation}(a) 
and (b)), the analytical prediction of Eq.~\eqref{psinusoidal} is in good agreement with the numerics. 
At high density (Fig.~\ref{fig:orientation}(c) and (d)), the theory provides a slight overestimation.
As expected, the average orientation increases with both the magnitude $v_a$ and angular frequency $\omega$ of the activity field. As can be seen in Fig.~\ref{fig:orientation}, the average orientation $\p(z)$ can attain significant values ($\approx 0.2$).
This is despite the fact that the rotational diffusion occurs on a time scale much smaller than the translational diffusion ($D_r/D_t = 25$). Since any activity field can be decomposed into a Fourier series, one can obtain the average orientation for a generic activity field. The only obvious limitation of the approach is that the theory predicts the linear order response and hence Eq.~\eqref{psinusoidal} is expected to be valid only for small activities.

\begin{figure}[t]
\includegraphics[width=0.99\columnwidth]{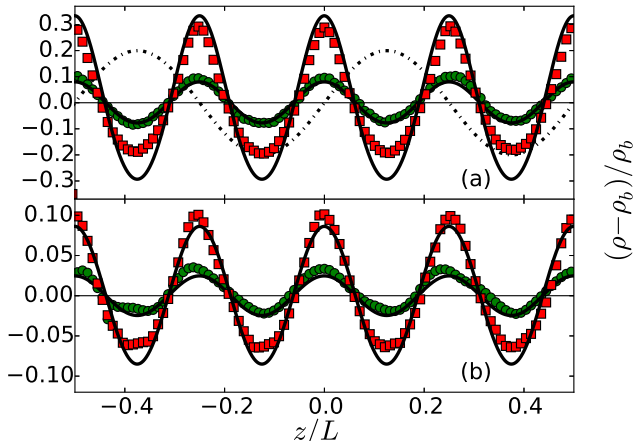} 
\caption{Relative change in density $(\rho-\rho_b)/\rho_b$ for $\rho_b = 0.2$ in (a) and $\rho_b = 0.6$ in (b). The dotted line shows a sinusoidal activity with $\omega = 2\omega_0$ with arbitrary amplitude. The circles correspond to $v_a = 10$ and the squares to $v_a = 20$. Particles accumulates at the nodes of the activity and the change in density is asymmetric. The thick lines correspond to the theoretical prediction of Eq.~\eqref{ddft} with Eq.~\eqref{psinusoidal} as input.}
\label{fig:density}
\end{figure}

We calculate the density by using Eqs.~\eqref{approximation} and \eqref{psinusoidal} as inputs to Eq.~\eqref{ddft}. In Fig.~\ref{fig:density} we plot the theoretical prediction together with the numerically measured relative change in density for two different bulk densities. There are three noteworthy features: (1) the density shows peaks at the nodes of the activity field, (2) the change in density becomes increasingly asymmetric with activity and (3) the relative change in density decreases with increasing initial bulk density. The theoretical prediction is consistent with these features. However, the theory predicts a more symmetric change in density than observed in simulations. 
Nevertheless, it is remarkable that a fundamentally second-order effect, i.e., the change in density, can be calculated theoretically to a good accuracy using a simple adiabatic DDFT approach. 
We note that in the absence of interparticle interactions Eq.~\eqref{ddft} can be easily solved 
in steady-state to yield a universal curve for the relative density change.
Given that the density change is as large as $30\%$ in Fig.~\ref{fig:density}(a) for $v_a = 20$, it is expected that nonlinear deviations appear in $\p(z)$. 
This is apparent in Fig.~\ref{fig:orientation}(a) which corresponds to the same set of parameters as Fig.~\ref{fig:density}(a). In this paper, we have not performed a systematic study of the range of validity of the linear response. Selective cases, with large activity, for which strong nonlinear deviations are observed in the system's reponse are presented in the supplementary material.

The underlying mechanism of orientation and density change can be understood qualitatively as follows: A particle with an orientation antiparallel to the activity gradient experiences slowing down as it moves in the direction of the gradient whereas a particle with orientation parallel to the gradient speeds up. This asymmetric influence of the activity, therefore, leads to accumulation of particles in proportion to the magnitude of the local gradient of activity.

To summarize our main findings: we have derived a formally exact expression (Eq.~\eqref{faverage}) 
for calculating averages in a system of interacting Brownian particles, subject to a 
position-dependent activity $v_0(\rr)$. From this we obtain the linear-response of the average orientation in Eq.~\eqref{pexpression} and 
identify the relevant time-correlation function as the self-part of the van Hove function.
We find that linear response provides an accurate account of the orientiation $\p(\rr)$ over a significant parameter range. Taking the analytical prediction for $\p(\rr)$ as an input to the dynamic density functional theory, we can obtain the spatial dependence of the density in good agreement with simulation data. Our approach is perfectly suited to obtain the time-dependent response of system subjected to a space-time dependent activity. It will be interesting to study the response $\p(\rr,t)$ and the corresponding time-evolution of density for a time- and space-dependent activity.


\begin{thebibliography}{30}
\expandafter\ifx\csname natexlab\endcsname\relax\def\natexlab#1{#1}\fi
\expandafter\ifx\csname bibnamefont\endcsname\relax
  \def\bibnamefont#1{#1}\fi
\expandafter\ifx\csname bibfnamefont\endcsname\relax
  \def\bibfnamefont#1{#1}\fi
\expandafter\ifx\csname citenamefont\endcsname\relax
  \def\citenamefont#1{#1}\fi
\expandafter\ifx\csname url\endcsname\relax
  \def\url#1{\texttt{#1}}\fi
\expandafter\ifx\csname urlprefix\endcsname\relax\def\urlprefix{URL }\fi
\providecommand{\bibinfo}[2]{#2}
\providecommand{\eprint}[2][]{\url{#2}}

\bibitem[{\citenamefont{Green}(1952)}]{green1952markoff}
\bibinfo{author}{\bibfnamefont{M.~S.} \bibnamefont{Green}},
  \bibinfo{journal}{The Journal of Chemical Physics}
  \textbf{\bibinfo{volume}{20}}, \bibinfo{pages}{1281} (\bibinfo{year}{1952}).

\bibitem[{\citenamefont{Kubo}(1957)}]{kubo1957statistical}
\bibinfo{author}{\bibfnamefont{R.}~\bibnamefont{Kubo}},
  \bibinfo{journal}{Journal of the Physical Society of Japan}
  \textbf{\bibinfo{volume}{12}}, \bibinfo{pages}{570} (\bibinfo{year}{1957}).

\bibitem[{\citenamefont{Sharma and Brader}(2016)}]{sharma2016communication}
\bibinfo{author}{\bibfnamefont{A.}~\bibnamefont{Sharma}} \bibnamefont{and}
  \bibinfo{author}{\bibfnamefont{J.}~\bibnamefont{Brader}},
  \bibinfo{journal}{The Journal of Chemical Physics}
  \textbf{\bibinfo{volume}{145}}, \bibinfo{pages}{161101}
  (\bibinfo{year}{2016}).

\bibitem[{\citenamefont{Cates and Tailleur}(2015)}]{cates2014motility}
\bibinfo{author}{\bibfnamefont{M.~E.} \bibnamefont{Cates}} \bibnamefont{and}
  \bibinfo{author}{\bibfnamefont{J.}~\bibnamefont{Tailleur}},
  \bibinfo{journal}{Annu. Rev. Condens. Matter Phys.}
  \textbf{\bibinfo{volume}{6}}, \bibinfo{pages}{219} (\bibinfo{year}{2015}).

\bibitem[{\citenamefont{Krinninger et~al.}(2016)\citenamefont{Krinninger,
  Schmidt, and Brader}}]{krinninger2016nonequilibrium}
\bibinfo{author}{\bibfnamefont{P.}~\bibnamefont{Krinninger}},
  \bibinfo{author}{\bibfnamefont{M.}~\bibnamefont{Schmidt}}, \bibnamefont{and}
  \bibinfo{author}{\bibfnamefont{J.~M.} \bibnamefont{Brader}},
  \bibinfo{journal}{Physical Review Letters} \textbf{\bibinfo{volume}{117}},
  \bibinfo{pages}{208003} (\bibinfo{year}{2016}).

\bibitem[{\citenamefont{Howse et~al.}(2007)\citenamefont{Howse, Jones, Ryan,
  Gough, Vafabakhsh, and Golestanian}}]{howse2007self}
\bibinfo{author}{\bibfnamefont{J.~R.} \bibnamefont{Howse}},
  \bibinfo{author}{\bibfnamefont{R.~A.} \bibnamefont{Jones}},
  \bibinfo{author}{\bibfnamefont{A.~J.} \bibnamefont{Ryan}},
  \bibinfo{author}{\bibfnamefont{T.}~\bibnamefont{Gough}},
  \bibinfo{author}{\bibfnamefont{R.}~\bibnamefont{Vafabakhsh}},
  \bibnamefont{and}
  \bibinfo{author}{\bibfnamefont{R.}~\bibnamefont{Golestanian}},
  \bibinfo{journal}{Physical review letters} \textbf{\bibinfo{volume}{99}},
  \bibinfo{pages}{048102} (\bibinfo{year}{2007}).

\bibitem[{\citenamefont{Jiang et~al.}(2010)\citenamefont{Jiang, Yoshinaga, and
  Sano}}]{jiang2010active}
\bibinfo{author}{\bibfnamefont{H.-R.} \bibnamefont{Jiang}},
  \bibinfo{author}{\bibfnamefont{N.}~\bibnamefont{Yoshinaga}},
  \bibnamefont{and} \bibinfo{author}{\bibfnamefont{M.}~\bibnamefont{Sano}},
  \bibinfo{journal}{Physical review letters} \textbf{\bibinfo{volume}{105}},
  \bibinfo{pages}{268302} (\bibinfo{year}{2010}).

\bibitem[{\citenamefont{Palacci et~al.}(2013)\citenamefont{Palacci, Sacanna,
  Steinberg, Pine, and Chaikin}}]{palacci2013living}
\bibinfo{author}{\bibfnamefont{J.}~\bibnamefont{Palacci}},
  \bibinfo{author}{\bibfnamefont{S.}~\bibnamefont{Sacanna}},
  \bibinfo{author}{\bibfnamefont{A.~P.} \bibnamefont{Steinberg}},
  \bibinfo{author}{\bibfnamefont{D.~J.} \bibnamefont{Pine}}, \bibnamefont{and}
  \bibinfo{author}{\bibfnamefont{P.~M.} \bibnamefont{Chaikin}},
  \bibinfo{journal}{Science} \textbf{\bibinfo{volume}{339}},
  \bibinfo{pages}{936} (\bibinfo{year}{2013}).

\bibitem[{\citenamefont{Gomez-Solano et~al.}(2016)\citenamefont{Gomez-Solano,
  Blokhuis, and Bechinger}}]{gomez2016dynamics}
\bibinfo{author}{\bibfnamefont{J.~R.} \bibnamefont{Gomez-Solano}},
  \bibinfo{author}{\bibfnamefont{A.}~\bibnamefont{Blokhuis}}, \bibnamefont{and}
  \bibinfo{author}{\bibfnamefont{C.}~\bibnamefont{Bechinger}},
  \bibinfo{journal}{Physical review letters} \textbf{\bibinfo{volume}{116}},
  \bibinfo{pages}{138301} (\bibinfo{year}{2016}).

\bibitem[{\citenamefont{Palacci et~al.}(2010)\citenamefont{Palacci,
  Ab{\'e}cassis, Cottin-Bizonne, Ybert, and Bocquet}}]{palacci2010colloidal}
\bibinfo{author}{\bibfnamefont{J.}~\bibnamefont{Palacci}},
  \bibinfo{author}{\bibfnamefont{B.}~\bibnamefont{Ab{\'e}cassis}},
  \bibinfo{author}{\bibfnamefont{C.}~\bibnamefont{Cottin-Bizonne}},
  \bibinfo{author}{\bibfnamefont{C.}~\bibnamefont{Ybert}}, \bibnamefont{and}
  \bibinfo{author}{\bibfnamefont{L.}~\bibnamefont{Bocquet}},
  \bibinfo{journal}{Physical review letters} \textbf{\bibinfo{volume}{104}},
  \bibinfo{pages}{138302} (\bibinfo{year}{2010}).

\bibitem[{\citenamefont{Saha et~al.}(2014)\citenamefont{Saha, Golestanian, and
  Ramaswamy}}]{saha2014clusters}
\bibinfo{author}{\bibfnamefont{S.}~\bibnamefont{Saha}},
  \bibinfo{author}{\bibfnamefont{R.}~\bibnamefont{Golestanian}},
  \bibnamefont{and}
  \bibinfo{author}{\bibfnamefont{S.}~\bibnamefont{Ramaswamy}},
  \bibinfo{journal}{Physical Review E} \textbf{\bibinfo{volume}{89}},
  \bibinfo{pages}{062316} (\bibinfo{year}{2014}).

\bibitem[{\citenamefont{Hong et~al.}(2007)\citenamefont{Hong, Blackman, Kopp,
  Sen, and Velegol}}]{hong2007chemotaxis}
\bibinfo{author}{\bibfnamefont{Y.}~\bibnamefont{Hong}},
  \bibinfo{author}{\bibfnamefont{N.~M.} \bibnamefont{Blackman}},
  \bibinfo{author}{\bibfnamefont{N.~D.} \bibnamefont{Kopp}},
  \bibinfo{author}{\bibfnamefont{A.}~\bibnamefont{Sen}}, \bibnamefont{and}
  \bibinfo{author}{\bibfnamefont{D.}~\bibnamefont{Velegol}},
  \bibinfo{journal}{Physical review letters} \textbf{\bibinfo{volume}{99}},
  \bibinfo{pages}{178103} (\bibinfo{year}{2007}).

\bibitem[{\citenamefont{Magiera and Brendel}(2015)}]{magiera2015trapping}
\bibinfo{author}{\bibfnamefont{M.~P.} \bibnamefont{Magiera}} \bibnamefont{and}
  \bibinfo{author}{\bibfnamefont{L.}~\bibnamefont{Brendel}},
  \bibinfo{journal}{Physical Review E} \textbf{\bibinfo{volume}{92}},
  \bibinfo{pages}{012304} (\bibinfo{year}{2015}).

\bibitem[{\citenamefont{Stenhammar et~al.}(2016)\citenamefont{Stenhammar,
  Wittkowski, Marenduzzo, and Cates}}]{stenhammar2016light}
\bibinfo{author}{\bibfnamefont{J.}~\bibnamefont{Stenhammar}},
  \bibinfo{author}{\bibfnamefont{R.}~\bibnamefont{Wittkowski}},
  \bibinfo{author}{\bibfnamefont{D.}~\bibnamefont{Marenduzzo}},
  \bibnamefont{and} \bibinfo{author}{\bibfnamefont{M.~E.} \bibnamefont{Cates}},
  \bibinfo{journal}{Science advances} \textbf{\bibinfo{volume}{2}},
  \bibinfo{pages}{e1501850} (\bibinfo{year}{2016}).

\bibitem[{\citenamefont{Berg et~al.}(1972)\citenamefont{Berg, Brown
  et~al.}}]{berg1972chemotaxis}
\bibinfo{author}{\bibfnamefont{H.~C.} \bibnamefont{Berg}},
  \bibinfo{author}{\bibfnamefont{D.~A.} \bibnamefont{Brown}},
  \bibnamefont{et~al.}, \bibinfo{journal}{Nature}
  \textbf{\bibinfo{volume}{239}}, \bibinfo{pages}{500} (\bibinfo{year}{1972}).

\bibitem[{\citenamefont{J{\'e}kely et~al.}(2008)\citenamefont{J{\'e}kely,
  Colombelli, Hausen, Guy, Stelzer, N{\'e}d{\'e}lec, and
  Arendt}}]{jekely2008mechanism}
\bibinfo{author}{\bibfnamefont{G.}~\bibnamefont{J{\'e}kely}},
  \bibinfo{author}{\bibfnamefont{J.}~\bibnamefont{Colombelli}},
  \bibinfo{author}{\bibfnamefont{H.}~\bibnamefont{Hausen}},
  \bibinfo{author}{\bibfnamefont{K.}~\bibnamefont{Guy}},
  \bibinfo{author}{\bibfnamefont{E.}~\bibnamefont{Stelzer}},
  \bibinfo{author}{\bibfnamefont{F.}~\bibnamefont{N{\'e}d{\'e}lec}},
  \bibnamefont{and} \bibinfo{author}{\bibfnamefont{D.}~\bibnamefont{Arendt}},
  \bibinfo{journal}{Nature} \textbf{\bibinfo{volume}{456}},
  \bibinfo{pages}{395} (\bibinfo{year}{2008}).

\bibitem[{\citenamefont{Hoff et~al.}(2009)\citenamefont{Hoff, van~der Horst,
  Nudel, and Hellingwerf}}]{hoff2009prokaryotic}
\bibinfo{author}{\bibfnamefont{W.~D.} \bibnamefont{Hoff}},
  \bibinfo{author}{\bibfnamefont{M.~A.} \bibnamefont{van~der Horst}},
  \bibinfo{author}{\bibfnamefont{C.~B.} \bibnamefont{Nudel}}, \bibnamefont{and}
  \bibinfo{author}{\bibfnamefont{K.~J.} \bibnamefont{Hellingwerf}},
  \bibinfo{journal}{Chemotaxis: Methods and Protocols} pp.
  \bibinfo{pages}{25--49} (\bibinfo{year}{2009}).

\bibitem[{\citenamefont{Jikeli et~al.}(2015)\citenamefont{Jikeli, Alvarez,
  Friedrich, Wilson, Pascal, Colin, Pichlo, Rennhack, Brenker, and
  Kaupp}}]{jikeli2015sperm}
\bibinfo{author}{\bibfnamefont{J.~F.} \bibnamefont{Jikeli}},
  \bibinfo{author}{\bibfnamefont{L.}~\bibnamefont{Alvarez}},
  \bibinfo{author}{\bibfnamefont{B.~M.} \bibnamefont{Friedrich}},
  \bibinfo{author}{\bibfnamefont{L.~G.} \bibnamefont{Wilson}},
  \bibinfo{author}{\bibfnamefont{R.}~\bibnamefont{Pascal}},
  \bibinfo{author}{\bibfnamefont{R.}~\bibnamefont{Colin}},
  \bibinfo{author}{\bibfnamefont{M.}~\bibnamefont{Pichlo}},
  \bibinfo{author}{\bibfnamefont{A.}~\bibnamefont{Rennhack}},
  \bibinfo{author}{\bibfnamefont{C.}~\bibnamefont{Brenker}}, \bibnamefont{and}
  \bibinfo{author}{\bibfnamefont{U.~B.} \bibnamefont{Kaupp}},
  \bibinfo{journal}{Nature communications} \textbf{\bibinfo{volume}{6}}
  (\bibinfo{year}{2015}).

\bibitem[{\citenamefont{Lozano et~al.}(2016)\citenamefont{Lozano, Ten~Hagen,
  L{\"o}wen, and Bechinger}}]{lozano2016phototaxis}
\bibinfo{author}{\bibfnamefont{C.}~\bibnamefont{Lozano}},
  \bibinfo{author}{\bibfnamefont{B.}~\bibnamefont{Ten~Hagen}},
  \bibinfo{author}{\bibfnamefont{H.}~\bibnamefont{L{\"o}wen}},
  \bibnamefont{and}
  \bibinfo{author}{\bibfnamefont{C.}~\bibnamefont{Bechinger}},
  \bibinfo{journal}{Nature communications} \textbf{\bibinfo{volume}{7}},
  \bibinfo{pages}{12828} (\bibinfo{year}{2016}).

\bibitem[{\citenamefont{Schweitzer et~al.}(1998)\citenamefont{Schweitzer,
  Ebeling, and Tilch}}]{schweitzer1998complex}
\bibinfo{author}{\bibfnamefont{F.}~\bibnamefont{Schweitzer}},
  \bibinfo{author}{\bibfnamefont{W.}~\bibnamefont{Ebeling}}, \bibnamefont{and}
  \bibinfo{author}{\bibfnamefont{B.}~\bibnamefont{Tilch}},
  \bibinfo{journal}{Physical Review Letters} \textbf{\bibinfo{volume}{80}},
  \bibinfo{pages}{5044} (\bibinfo{year}{1998}).

\bibitem[{\citenamefont{Enculescu and Stark}(2011)}]{enculescu2011active}
\bibinfo{author}{\bibfnamefont{M.}~\bibnamefont{Enculescu}} \bibnamefont{and}
  \bibinfo{author}{\bibfnamefont{H.}~\bibnamefont{Stark}},
  \bibinfo{journal}{Physical review letters} \textbf{\bibinfo{volume}{107}},
  \bibinfo{pages}{058301} (\bibinfo{year}{2011}).

\bibitem[{\citenamefont{Morriss and Evans}(2007)}]{morriss2007statistical}
\bibinfo{author}{\bibfnamefont{G.~P.} \bibnamefont{Morriss}} \bibnamefont{and}
  \bibinfo{author}{\bibfnamefont{D.~J.} \bibnamefont{Evans}},
  \emph{\bibinfo{title}{Statistical Mechanics of Nonequilbrium Liquids}}
  (\bibinfo{publisher}{ANU Press}, \bibinfo{year}{2007}).

\bibitem[{\citenamefont{Fuchs and Cates}(2002)}]{fuchs2002theory}
\bibinfo{author}{\bibfnamefont{M.}~\bibnamefont{Fuchs}} \bibnamefont{and}
  \bibinfo{author}{\bibfnamefont{M.~E.} \bibnamefont{Cates}},
  \bibinfo{journal}{Physical review letters} \textbf{\bibinfo{volume}{89}},
  \bibinfo{pages}{248304} (\bibinfo{year}{2002}).

\bibitem[{\citenamefont{Fuchs and Cates}(2005)}]{fuchs2005integration}
\bibinfo{author}{\bibfnamefont{M.}~\bibnamefont{Fuchs}} \bibnamefont{and}
  \bibinfo{author}{\bibfnamefont{M.~E.} \bibnamefont{Cates}},
  \bibinfo{journal}{Journal of Physics: Condensed Matter}
  \textbf{\bibinfo{volume}{17}}, \bibinfo{pages}{S1681} (\bibinfo{year}{2005}).

\bibitem[{\citenamefont{Brader et~al.}(2012)\citenamefont{Brader, Cates, and
  Fuchs}}]{brader2012first}
\bibinfo{author}{\bibfnamefont{J.~M.} \bibnamefont{Brader}},
  \bibinfo{author}{\bibfnamefont{M.~E.} \bibnamefont{Cates}}, \bibnamefont{and}
  \bibinfo{author}{\bibfnamefont{M.}~\bibnamefont{Fuchs}},
  \bibinfo{journal}{Physical Review E} \textbf{\bibinfo{volume}{86}},
  \bibinfo{pages}{021403} (\bibinfo{year}{2012}).

\bibitem[{\citenamefont{Hansen and McDonald}(1990)}]{hansen1990theory}
\bibinfo{author}{\bibfnamefont{J.-P.} \bibnamefont{Hansen}} \bibnamefont{and}
  \bibinfo{author}{\bibfnamefont{I.~R.} \bibnamefont{McDonald}},
  \emph{\bibinfo{title}{Theory of simple liquids}}
  (\bibinfo{publisher}{Elsevier}, \bibinfo{year}{1990}).

\bibitem[{\citenamefont{Gardiner et~al.}(1985)}]{gardiner1985handbook}
\bibinfo{author}{\bibfnamefont{C.~W.} \bibnamefont{Gardiner}}
  \bibnamefont{et~al.}, \emph{\bibinfo{title}{Handbook of stochastic methods}},
  vol.~\bibinfo{volume}{3} (\bibinfo{publisher}{Springer Berlin},
  \bibinfo{year}{1985}).

\bibitem[{\citenamefont{Morse et~al.}(1953)\citenamefont{Morse, Feshbach
  et~al.}}]{morse1953methods}
\bibinfo{author}{\bibfnamefont{P.~M.} \bibnamefont{Morse}},
  \bibinfo{author}{\bibfnamefont{H.}~\bibnamefont{Feshbach}},
  \bibnamefont{et~al.}, \emph{\bibinfo{title}{Methods of theoretical physics}},
  vol.~\bibinfo{volume}{1} (\bibinfo{publisher}{McGraw-Hill New York},
  \bibinfo{year}{1953}).

\bibitem[{\citenamefont{Bickel et~al.}(2014)\citenamefont{Bickel, Zecua, and
  W{\"u}rger}}]{bickel2014polarization}
\bibinfo{author}{\bibfnamefont{T.}~\bibnamefont{Bickel}},
  \bibinfo{author}{\bibfnamefont{G.}~\bibnamefont{Zecua}}, \bibnamefont{and}
  \bibinfo{author}{\bibfnamefont{A.}~\bibnamefont{W{\"u}rger}},
  \bibinfo{journal}{Physical Review E} \textbf{\bibinfo{volume}{89}},
  \bibinfo{pages}{050303} (\bibinfo{year}{2014}).

\bibitem[{\citenamefont{Hopkins et~al.}(2010)\citenamefont{Hopkins, Fortini,
  Archer, and Schmidt}}]{hopkins2010van}
\bibinfo{author}{\bibfnamefont{P.}~\bibnamefont{Hopkins}},
  \bibinfo{author}{\bibfnamefont{A.}~\bibnamefont{Fortini}},
  \bibinfo{author}{\bibfnamefont{A.~J.} \bibnamefont{Archer}},
  \bibnamefont{and} \bibinfo{author}{\bibfnamefont{M.}~\bibnamefont{Schmidt}},
  \bibinfo{journal}{The Journal of chemical physics}
  \textbf{\bibinfo{volume}{133}}, \bibinfo{pages}{224505}
  (\bibinfo{year}{2010}).

\end{thebibliography}

\end{document}